\documentclass[aps,pla,groupedaddress,showpacs]{revtex4}
\usepackage{bbm}
\usepackage{amsfonts}
\usepackage{amsmath}
\usepackage{enumerate}
\usepackage{graphicx}
\begin{document}
\title{Applications of Polynomial Algebras to 2-Dimensional Deformed Oscillators}
\author{Ci Song}
\email[Email:]{cisong@mail.nankai.edu.cn} \affiliation{Theoretical
Physics Division, Chern Institute of Mathematics, Nankai University,
Tianjin 300071, People's Republic of China}

\author{Fu-Lin Zhang}
\email[Email:]{flzhang@mail.nankai.edu.cn} \affiliation{Theoretical
Physics Division, Chern Institute of Mathematics, Nankai University,
Tianjin 300071, People's Republic of China}

\author{Jing-Ling Chen}
\email[Email:]{chenjl@nankai.edu.cn} \affiliation{Theoretical
Physics Division, Chern Institute of Mathematics, Nankai University,
Tianjin 300071, People's Republic of China}

\date{\today}

\begin{abstract}
The polynomial algebra is a deformed su(2) algebra. Here, we use
polynomial algebra as a method to solve a series of deformed
oscillators. Meanwhile, we find a series of physics systems
corresponding with polynomial algebra with different maximal order.
\end{abstract}

\pacs{02.20.-a; 03.65.Fd; 03.65.Ge; 03.65.-w}

\maketitle

\section{Introduction}

The idea of using physical systems symmetries to study degenerate
energy levels has been adopted since the early days of quantum
mechanics. So ladder operators which connect all the eigen-states
with a given energy lead a good method to solve this problem. For
linear systems, such as Hydrogen atom and isotropic harmonic
oscillator, Lie algebra can work out these problems well. Generally,
the $N$-dimensional  hydrogen atom has the so($N+1$) and the
oscillator has the su($N$) symmetry.

Afterwards, Higgs \cite{Higgs} and Leemon \cite{Leemon} introduced a
generalization of the hydrogen atom and isotropic harmonic
oscillator in a space with constant curvature. In Higgs' literature
\cite{Higgs}, he constructed a new algebra isomoriphic to so($3$)
and su($2$) to describe the symmetry of hydrogen atom and isotropic
harmonic oscillator on 2-dimensional sphere and this new algebra is
called Higgs algebra which is also used in two-body
Calogero-Sutherland model \cite{CSM} and Karassiov-Klimov model
\cite{KKM}. Then, additional examples, like the Fokas-Lagerstrom
potential \cite{FLP}, the Smorodinsky-Winternitz potential
\cite{SW1}, and the Holt potential \cite{Holt}, were finally solved
by Dennis Bonatsos et al \cite{Higgs2} in the method of ladder
operators.

The polynomial algebra \cite{D.Ruan} is a deformation of normal
angular algebra su($2$), which owns three generators $J_0, J_+$ and
$J_-$. However, the commutative relation of $J_+$ and $J_-$ appears
the polynomial of $J_0$. su($2$) and Higgs algebra are both special
cases of polynomial algebra. It can be represented as
$\mathfrak{J}^{(\Omega)}$, where $\Omega$ is a positive integer
which expresses the highest power of the polynomial. The generators
$J_0, J_+$ and $J_-$ of $\mathfrak{J}^{(\Omega)}$ satisfy
\begin{eqnarray}\label{eq:comm_DR_J0+-}
[J_0,J_\pm]=\pm J_\pm,\quad[J_+,J_-]=P(J_0),
\end{eqnarray}
and its Casimir operator can be written as
\begin{eqnarray}\label{eq:Oper_DR_Casimir}
C^{(\Omega)}=\{J_+,J_-\}+\sum_{i=0}^{\Omega+1}\alpha_iJ_0^i.
\end{eqnarray}
Here, in this paper, we expand the Fokas-Lagerstrom potential and
the Holt potential to the oscillator's frequency satisfying
$\omega_1:\omega_2=l_1:l_2$ which is integer ratio. Thus, with this
result, we can easily get Bonatsos' result.

\section{Polynomial algebra method}

For a 2-dimensional physical system exhibiting dynamical symmetry,
we can find a set of operators $J_0, J_+$ and $J_-$ which
communicate with the Hamiltonian of system and satisfy
(\ref{eq:comm_DR_J0+-}) as ladder operators.

Firstly, we assume the dimension of representation of this system is
finite. So there must be an upper bound
$\left|\overline{\mathbbm{m}}\right>$ and a lower bound
$\left|\underline{\mathbbm{m}}\right>$ in each degenerate energy
level.

Meanwhile, because of (\ref{eq:comm_DR_J0+-}), it is easy to see
$[J_+J_-,J_0]=[J_-J_+,J_0]=0$. So we know $J_+J_-$ and $J_-J_+$ must
be the function of $J_0$ and $H$
\begin{eqnarray}\label{eq:Oper_J+J-_phi}
J_+J_-=\phi(J_0,H),\quad J_-J_+=\phi(J_0+1,H)
\end{eqnarray}
Thus, we use them to act on $\left|\overline{\mathbbm{m}}\right>$
and $\left|\underline{\mathbbm{m}}\right>$ respectively. We get
equations
\begin{eqnarray}
J_+J_-\left|\underline{\mathbbm{m}}\right>=\phi(\underline{\mathbbm{m}},E)\left|\underline{\mathbbm{m}}\right>=0,\quad
J_-J_+\left|\overline{\mathbbm{m}}\right>=\phi(\overline{\mathbbm{m}}+1,E)\left|\overline{\mathbbm{m}}\right>=0.
\end{eqnarray}
In both equations, we can omit energy $E$ and require
$\overline{\mathbbm{m}}-\underline{\mathbbm{m}}=n$ is integer.
Finally, we could omit part of results which cause energy $E$ goes
to negative infinity when $n$ goes to positive infinity. Then, we
finally get the energy level and degenerate degree.

\section{Using polynomial algebra in 2-Dimensional Deformed Oscillators}
\subsection{2-Dimension isotropic harmonic oscillator}
Firstly, we use 2-D isotropic harmonic oscillator as an example. Its
Hamiltonian can be written as
\begin{eqnarray}\label{eq:hamil_iHO}
H=\frac{p_1^2+p_2^2}{2m}+\frac{1}{2}m\omega^2(x_1^2+x_2^2).
\end{eqnarray}
If we write operators
\begin{equation}\label{eq:Oper_LiftLow}
a_i=\sqrt{\frac{m\omega_i}{2\hbar}}x_i+i\frac{p_i}{\sqrt{2m\omega_i\hbar}},\quad
a_i^{\dagger}=\sqrt{\frac{m\omega_i}{2\hbar}}x_i-i\frac{p_i}{\sqrt{2m\omega_i\hbar}}\quad(i=1,2)
\end{equation}
and
\begin{equation}\label{eq:Oper_Num}
N_i=a_i^{\dagger}a_i=\frac{1}{\hbar\omega_i}(\frac{p_i^2}{2m}+\frac{1}{2}m\omega_i^2x_i^2)-\frac{1}{2}\quad(i=1,2),
\end{equation}
We can rewrite the Hamiltonian as the following form
\begin{equation}\label{eq:Repre_iHO_hamil_Num}
H=\hbar\omega(N_1+N_2+1)
\end{equation}
\begin{enumerate}
\item Normal method

Usually it is solved by second order tensors \cite{HO_2Tensor}.
\begin{eqnarray}\label{eq:Oper_iHO_2Tensor}
S_0=\frac{1}{2}(N_1-N_2),\quad S_+=a_1^\dagger a_2,\quad
S_-=a_1a_2^\dagger.
\end{eqnarray}
Their Casimir operator $C$ can be write as
\begin{eqnarray}\label{eq:Oper_iHO_2Tensor_Casimir}
C=\frac{1}{2}\{S_+,S_-\}+S_0^2,
\end{eqnarray}
and energy level can be solved as
\begin{eqnarray}\label{eq:EL_iHO}
E_n=\hbar\omega(n+1),\quad n=n_1+n_2,\quad n_1,n_2=0,1,2,\cdots
\end{eqnarray}
where $n=0,1,2,\cdots$, and there are $n+1$ degenerate eigenstates
for each energy level $E_n$.
\item Polynomial algebra method

If we use new operators
\begin{equation}\label{eq:Oper_HO_J0+-}
J_0=\frac{1}{4}(N_1-N_2),\quad J_+=(a_1^\dagger)^2(a_2)^2,\quad
J_-=(a_1)^2(a_2^\dagger)^2.
\end{equation}
We could find their communicative relations
\begin{equation}\label{eq:comm_HO_J0+-}
[J_0,J_+]=J_+,\quad[J_0,J_-]=-J_-,
\end{equation}
and
\begin{eqnarray}\label{eq:comm_HO_J+J-}
[J_+,J_-]=4\left(\frac{H^2}{\hbar^2\omega^2}-3\right)J_0-64J_0^3.
\end{eqnarray}
From the communicative relation, it is obvious that $J_+,\ J_-$ and
$J_0$ satisfies Higgs algebra relation\cite{Higgs}, which the
maximal order of $J_0$ in $[J_+,J_-]$ is $3$. Meanwhile, we can get
their Casimir operator
\begin{eqnarray}\label{eq:Oper_HO_Casimir}
C=\frac{1}{8}\left(\frac{H}{\hbar\omega}\right)^4-\frac{5}{4}\left(\frac{H}{\hbar\omega}\right)^2+\frac{9}{8},
\end{eqnarray}
and energy level
\begin{subequations}
\begin{eqnarray}
\label{eq:HO_energy1}E_{1,n}&=&\hbar\omega(2n+1)\\
\label{eq:HO_energy2}E_{2,n}&=&\hbar\omega(2n+2)
\end{eqnarray}
\end{subequations}
\begin{figure}
  \centering
  \includegraphics[width=2in]{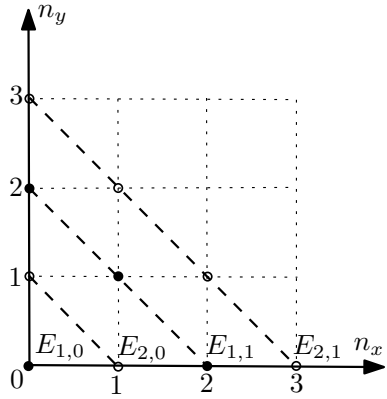}\\
  \caption{As shown in the figure, the solid point is represented the energy state described by Eq.(\ref{eq:HO_energy1})
   and the hollow point is represented the energy state described by Eq.(\ref{eq:HO_energy2}).
   The dashed line shows the degenerate eigenstates.}\label{Fig:HO}
\end{figure}

where $n=0,1,2,\cdots$, and there are $2n+i$ degenerate eigenstates
for each energy level $E_{in}$, $i=1,2$. As shown in
Fig\ref{Fig:HO}, the solid point is represented the energy state
described by Eq.(\ref{eq:HO_energy1}) and the hollow point is
represented the energy state described by Eq.(\ref{eq:HO_energy2}).
\end{enumerate}
Comparing above two methods, we can see the polynomial algebra can
also give all the energy level for the system. More exciting, it
could be used for other deformed oscillator or non-linear potential.

\subsection{2-Dimensional anisotropic harmonic oscillator}
The Hamiltonian of 2-D anisotropic harmonic oscillator can be
written as
\begin{eqnarray}\label{eq:hamiltonian_LSR}
H=\frac{p_1^2+p_2^2}{2m}+\frac{1}{2}m(\omega_1^2x_1^2+\omega_2^2x_2^2)=(N_1+\frac{1}{2})\hbar\omega_1+(N_2+\frac{1}{2})\hbar\omega_2.
\end{eqnarray}
If $\omega_1:\omega_2=l_1:l_2$ is integer ratio, we can write
$\omega_1=l_1\omega_0,\omega_2=l_2\omega_0$ and construct new
operators
\begin{eqnarray}\label{eq:Oper_LSR_vecJ}
J_0=\frac{1}{2}\left(\frac{N_1}{l_2}-\frac{N_2}{l_1}\right),\quad
J_+=(a_1^{\dagger})^{l_2}(a_2)^{l_1},\quad
J_-=(a_1)^{l_2}(a_2^{\dagger})^{l_1}.
\end{eqnarray}
We could find their communicative relations
\begin{equation}\label{eq:comm_LSR_J0+-}
[J_0,J_+]=J_+,\quad[J_0,J_-]=-J_-
\end{equation}
and
\begin{eqnarray}\label{eq:comm_LSR_J+J-}
[J_+,J_-]&=&\prod_{i=1}^{l_2}\left(\frac{2H-\hbar\omega_2}{4\hbar\omega_1}+l_2J_0-i+\frac{3}{4}\right)\cdot\prod_{j=1}^{l_1}\left(\frac{2H-\hbar\omega_1}{4\hbar\omega_2}-l_1J_0+j-\frac{1}{4}\right)\nonumber\\
&&-\prod_{i=1}^{l_2}\left(\frac{2H-\hbar\omega_2}{4\hbar\omega_1}+l_2J_0+i-\frac{1}{4}\right)\cdot\prod_{j=1}^{l_1}\left(\frac{2H-\hbar\omega_1}{4\hbar\omega_2}-l_1J_0-j+\frac{3}{4}\right).\nonumber\\
&&
\end{eqnarray}
which the maximal order of $J_0$ in $[J_+,J_-]$ is $l_1+l_2-1$
corresponding to the polynomial algebras with $l_1+l_2-1$ order. We
can solve their Casimir operator
\begin{eqnarray}\label{eq:Oper_LSR_Casimir}
C&=&\prod_{i=1}^{l_2}\left(\frac{2H-\hbar\omega_2}{4\hbar\omega_1}-i+\frac{3}{4}\right)\cdot\prod_{j=1}^{l_1}\left(\frac{2H-\hbar\omega_1}{4\hbar\omega_2}+j-\frac{1}{4}\right)\\
&&+\prod_{i=1}^{l_2}\left(\frac{2H-\hbar\omega_2}{4\hbar\omega_1}+i-\frac{1}{4}\right)\cdot\prod_{j=1}^{l_1}\left(\frac{2H-\hbar\omega_1}{4\hbar\omega_2}-j+\frac{3}{4}\right).\nonumber
\end{eqnarray}
and energy level
\begin{eqnarray}\label{eq:LSR_energy_level}
E_{i,j;n}=\hbar\omega_1(i-\frac{1}{2})+\hbar\omega_2(j-\frac{1}{2})+\hbar\frac{\omega_1\omega_2}{\omega_0}n
\end{eqnarray}
where $n=0,1,2,\cdots$, and there are $n+1$ degenerate eigenstates
for each energy level $E_{i,j;n}$, $i=1,\cdots,l_2$,
$j=1,\cdots,l_1$, which different $i$ and $j$ numbers show different
formulae for the energy levels.

\begin{figure}
  \centering
  \includegraphics[width=2in]{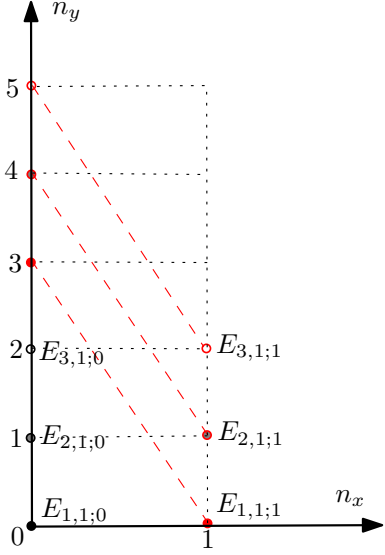}\\
  \caption{As shown in the figure(color online), the solid point is represented the energy state described by Eq.(\ref{eq:HO_3_1_energy_level_1});
   the grey point is represented the energy state described by Eq.(\ref{eq:HO_3_1_energy_level_2}); the hollow point is represented the energy state described by Eq.(\ref{eq:HO_3_1_energy_level_3}).
   The red dashed line shows the degenerate eigenstates when $n=1$.}\label{Fig:HO_3_1}
\end{figure}

When $l_1:l_2=3:1$, it could be viewed as Fokas-Lagerstorm
potential, which is taken as an example here. For Fokas-Lagerstorm
potential, we can calculate its communicative relation as
\begin{eqnarray}
[J_+,J_-]&=&\frac{1}{64\hbar^3\omega_1\omega_2^3}\left(-8H^3(\omega_1-9\omega_2)+12\hbar
H^2(\omega_1^2-4\omega_1\omega_2+3\omega_2^2)-\right.\nonumber\\
&&\left.2\hbar^2H(3\omega_1^3+3\omega_1^2\omega_2+77\omega_1\omega_2^2-51\omega_2^3)+\hbar^3(\omega_1^4+6\omega_1^3\omega_2+68\omega_1^2\omega_2^2-6\omega_1\omega_2^3-69\omega_2^4)\right)\nonumber\\
&&+\frac{3\left(12H^2(\omega_1-3\omega_2)-12\hbar
H\omega_1(\omega_1-\omega_2)+\hbar^2(3\omega_1^3+3\omega_1^2\omega_2+41\omega_1\omega_2^2+9\omega_2^3)\right)}{8\hbar^2\omega_1\omega_2^2}J_0\nonumber\\
&&+\frac{81\left(-2H\omega_1+\hbar\omega_1^2+2H\omega_2-\hbar\omega_2^2\right)}{4\hbar\omega_1\omega_2}J_0^2+108J_0^3
\end{eqnarray}
which the maximal order of $J_0$ in $[J_+,J_-]$ is $3$ corresponding
to the polynomial algebras with $3$ order. The Casimir operator can
be expressed as
\begin{eqnarray}
C&=&-\frac{1}{128\hbar^4\omega_1\omega_2^3}\left(-16H^4+16\hbar
H^3(\omega_1-\omega_2)+16\hbar^2
H^2(9\omega_1-23\omega_2)\omega_2\right.\nonumber\\
&&\left.-4\hbar^3H(\omega_1^3+33\omega_1^2\omega_2-33\omega_1\omega_2^2-\omega_2^3)+\hbar^4(\omega_1^4+32\omega_1^3\omega_2+26\omega_1^2\omega_2^2+88\omega_1\omega_2^3+93\omega_2^4)\right)\nonumber\\
\end{eqnarray}
the energy level calculated by polynomial algebra can array as
follows
\begin{subequations}\label{eq:HO_3_1_energy_level}
\begin{eqnarray}
\label{eq:HO_3_1_energy_level_1}E_{1,1;n}&=&\frac{1}{2}\hbar\omega_1+\frac{1}{2}\hbar\omega_2+\hbar\frac{\omega_1\omega_2}{\omega_0}n,\\
\label{eq:HO_3_1_energy_level_2}E_{2,1;n}&=&\frac{3}{2}\hbar\omega_1+\frac{1}{2}\hbar\omega_2\hbar\frac{\omega_1\omega_2}{\omega_0}n,\\
\label{eq:HO_3_1_energy_level_3}E_{3,1;n}&=&\frac{5}{2}\hbar\omega_1+\frac{1}{2}\hbar\omega_2+\hbar\frac{\omega_1\omega_2}{\omega_0}n.
\end{eqnarray}
\end{subequations}
For $l_1=3$ and $l_2=1$, it is clearly that the energy levels have
three formula forms, which the number of energy level formulae
equals to $l_1\times l_2$. As shown in Fig\ref{Fig:HO_3_1}, we have
known that the number of degenerate eigenstates equals to $n+1$ for
each energy level formula.

\subsection{2-Dimensional anisotropic harmonic oscillator with Smorodinsky-Winternitz potential}
The Hamiltonian of 2-Dimensional anisotropic harmonic oscillator
with Smorodinsky-Winternitz potential system can be written as
\begin{eqnarray}\label{eq:hamiltonian_unsymmetry}
H=\frac{p_1^2+p_2^2}{2m}+\frac{1}{2}m\omega_1^2x_1^2+\frac{1}{2}m\omega_2^2x_2^2+\frac{\kappa}{2x_2^2}
\end{eqnarray}
where $V_I=\frac{\kappa}{2x_2^2}$ is hard to deal with. We can
construct operators
\begin{subequations}\label{eq:Oper_LiftLow_New}
\begin{eqnarray}
A_1&=&a_1^2,\\
A_1^{\dagger}&=&(a_1^{\dagger})^2\\
A_2&=&a_2^2-\frac{V_I}{\hbar\omega_2},\\
A_2^{\dagger}&=&(a_2^{\dagger})^2-\frac{V_I}{\hbar\omega_2},
\end{eqnarray}
\end{subequations}
and rewrite the Hamiltonian as
\begin{eqnarray}\label{eq:Repre_hamiltonian_unsymmetry_Num}
H=H_1+H_2,\quad H_1=(N_1+\frac{1}{2})\hbar\omega_1,\quad
H_2=(N_2+\frac{1}{2})\hbar\omega_2+V_I
\end{eqnarray}
They satisfy communicative relations as
\begin{eqnarray}\label{eq:comm_H_LiftLow_Square}
\left[H_i,A_j\right]=-2\hbar\omega_iA_i\delta_{ij},\quad
\left[H_i,A_j^{\dagger}\right]=2\hbar\omega_iA_i^{\dagger}\delta_{ij},\quad
\left[A_i,A_j^{\dagger}\right]=\frac{4}{\hbar\omega_i}H_i\delta_{ij}.
\end{eqnarray}
So, for total Hamiltonian (\ref{eq:hamiltonian_unsymmetry}), we have
\begin{equation}\label{eq:Relation_hamil_constant_1}
H(A_1)^{l_2}(A_2^{\dagger})^{l_1}=(A_1)^{l_2}(A_2^{\dagger})^{l_1}(H+2{l_1}\hbar\omega_2-2{l_2}\hbar\omega_1)
\end{equation}
which means that, if $\omega_1=l_1\omega_0,\ \omega_2=l_2\omega_0$
is integer ratio, we have $[H,(A_1)^{l_2}(A_2^{\dagger})^{l_1}]=0$.
So we can construct the ladder operators
\begin{eqnarray}\label{eq:new_SWP_J0-+}
J_0=\frac{1}{2(l_1+l_2)\hbar}\left(\frac{H_1}{\omega_1}-\frac{H_2}{\omega_2}\right),\quad
J_+=(A_1^\dagger)^{l_2}A_2^{l_1},\quad
J_-=A_1^{l_2}(A_2^\dagger)^{l_1}
\end{eqnarray}
We could find their communicative relations
\begin{eqnarray}\label{eq:comm_new_SW-potential_J0-+}
[J_0,J_+]=J_+,\quad[J_0,J_-]=-J_-
\end{eqnarray}
and
\begin{eqnarray}\label{eq:comm_new_SWP_J+-}
[J_+,J_-]&=&\prod_{i=0}^{l_2-1}\left(\frac{H}{\hbar(\omega_1+\omega_2)}+2l_2J_0+(2i-\frac{1}{2})\right)\cdot\left(\frac{H}{\hbar(\omega_1+\omega_2)}+2l_2J_0+(2i-\frac{3}{2})\right)\cdot\nonumber\\
&&\prod_{j=0}^{l_1-1}\left(\frac{H}{\hbar(\omega_1+\omega_2)}-2l_1J_0-(2j-1)-\frac{1}{2}\sqrt{\frac{4m\kappa}{\hbar^2}+1}\right)\cdot\nonumber\\
&&\left(\frac{H}{\hbar(\omega_1+\omega_2)}-2l_1J_0-(2j-1)+\frac{1}{2}\sqrt{\frac{4m\kappa}{\hbar^2}+1}\right)\nonumber\\
&&-\prod_{i=0}^{l_2-1}\left(\frac{H}{\hbar(\omega_1+\omega_2)}+2l_2J_0-(2i-\frac{3}{2})\right)\cdot\left(\frac{H}{\hbar(\omega_1+\omega_2)}+2l_2J_0-(2i-\frac{1}{2})\right)\cdot\nonumber\\
&&\prod_{j=0}^{l_1-1}\left(\frac{H}{\hbar(\omega_1+\omega_2)}-2l_1J_0+(2j-1)-\frac{1}{2}\sqrt{\frac{4m\kappa}{\hbar^2}+1}\right)\cdot\nonumber\\
&&\left(\frac{H}{\hbar(\omega_1+\omega_2)}-2l_1J_0+(2j-1)+\frac{1}{2}\sqrt{\frac{4m\kappa}{\hbar^2}+1}\right)
\end{eqnarray}
which the maximal order of $J_0$ in $[J_+,J_-]$ is $2(l_1+l_2)-1$
corresponding to the polynomial algebras with $2(l_1+l_2)-1$ order.
We can solve their Casimir operator
\begin{eqnarray}\label{eq:Oper_new_SWP_Casimir}
C&=&\prod_{i=0}^{{l_2}-1}\left(\left(\frac{H}{\hbar(\omega_1+\omega_2)}\right)^2+2\frac{H(2i-1)}{\hbar(\omega_1+\omega_2)}+4i(i-1)+\frac{3}{4}\right)\cdot\\
&&\prod_{j=0}^{{l_1}-1}\left(\left(\frac{H}{\hbar(\omega_1+\omega_2)}\right)^2-\frac{2H(2j-1)}{\hbar(\omega_1+\omega_2)}+4j(j-1)+\frac{3}{4}-\frac{m\kappa}{\hbar^2}\right)\nonumber\\
&&+\prod_{i=0}^{{l_2}-1}\left(\left(\frac{H}{\hbar(\omega_1+\omega_2)}\right)^2-\frac{2H(2i-1)}{\hbar(\omega_1+\omega_2)}+4i(i-1)+\frac{3}{4}\right)\cdot\nonumber\\
&&\prod_{j=0}^{{l_1}-1}\left(\left(\frac{H}{\hbar(\omega_1+\omega_2)}\right)^2+\frac{2H(2j-1)}{\hbar(\omega_1+\omega_2)}+4j(j-1)+\frac{3}{4}-\frac{m\kappa}{\hbar^2}\right)\nonumber
\end{eqnarray}
and energy level
\begin{subequations}
\begin{eqnarray}
E_{(1)i,j;n}=2\frac{\omega_1\omega_2}{\omega_0}n-\hbar\omega_1(2i-\frac{1}{2})-\hbar\omega_2(2j-1)-\frac{\omega_2}{2}\sqrt{4m\kappa+\hbar^2};\\
E_{(2)i,j;n}=2\frac{\omega_1\omega_2}{\omega_0}n-\hbar\omega_1(2i-\frac{1}{2})-\hbar\omega_2(2j-1)+\frac{\omega_2}{2}\sqrt{4m\kappa+\hbar^2};\\
E_{(3)i,j;n}=2\frac{\omega_1\omega_2}{\omega_0}n-\hbar\omega_1(2i-\frac{3}{2})-\hbar\omega_2(2j-1)-\frac{\omega_2}{2}\sqrt{4m\kappa+\hbar^2};\\
E_{(4)i,j;n}=2\frac{\omega_1\omega_2}{\omega_0}n-\hbar\omega_1(2i-\frac{3}{2})-\hbar\omega_2(2j-1)+\frac{\omega_2}{2}\sqrt{4m\kappa+\hbar^2}.
\end{eqnarray}
\end{subequations}
where $-\frac{\hbar^2}{4m}<\kappa<\frac{3\hbar^2}{4m}$,
$n=0,1,2,\cdots$ and there are $n+1$ degenerate eigenstates for each
energy level $E_{(s)nij}$, $s=1,2,3,4$, $i=0,\cdots,l_2-1,\
j=0,\cdots,l_1-1$. When $l_1:l_2=1:1$, it could be viewed as the
Smorodinsky-Winternitz potential; when $l_1:l_2=1:2$, it could be
viewed as the Holt potential.

\section{Discussion}

In this paper, we solve arbitrary integer ratio, $l_1:l_2$, between
two frequencies of 2-dimensional harmonic oscillator. The deformed
oscillators could be solved by polynomial algebras. Meanwhile,
oscillators with arbitrary integer ratio frequencies are also real
physical model. Actually, with ladder operators, the physical model
with equal energy interval can be solved by polynomial algebras.
With this practice of 2-dimensional system, we could try to solve
3-dimensional system with expanding su($3$) or so($4$) to their
non-linear form.

\section*{Acknowledge}
We thank Bo Fu for his helpful discussion and checking the
manuscript carefully. This work is supported in part by NSF of China
(Grants No. 10975075), Program for New Century Ex- cellent Talents
in University, and the Project-sponsored 5 by SRF for ROCS, SEM.

\end{document}